\documentclass[doublecol]{epl2}%
\usepackage{amssymb}
\usepackage{amsmath}
\usepackage{amsfonts}
\usepackage{graphicx}%
\providecommand{\U}[1]{\protect\rule{.1in}{.1in}}
\shorttitle{Nuclear polariton lineshape}
\shortauthor{A.I. Rykov \etal}
\institute{
\inst{1}School of Engineering, The University of Tokyo, Hongo 7-3-1, Bunkyo-ku, Tokyo 113-8656, Japan\\
\inst{2}Siberian Synchrotron Radiation Center, Lavrentieva 11, 630090, Novosibirsk, Russia
}
\abstract{In synchrotron M\"ossbauer spectroscopy, the nuclear exciton polariton manifests itself in the lineshape of the spectra of nuclear forward scattering (NFS) Fourier-transformed from time domain to frequency domain. This lineshape is generally described by the convolution of two intensity factors. One of them is Lorentzian related to free decay. We derived the expressions for the second factor related to Frenkel exciton polariton effects at propagation of synchrotron radiation in M\"ossbauer media. Parameters of this Frenkelian shape depend on the spatial configuration of M\"ossbauer media. In a layer of uniform thickness, this factor is found to be a simple hypergeometric function. Next, we consider the particles spread over a 2D surface or diluted in non-M\"ossbauer media to exclude an overlap of ray shadows by different particles. Deconvolving the purely polaritonic component of linewidths is suggested as a simple procedure sharpening  the  experimental NFS spectra in frequency domain. The lineshapes in these sharpened spectra are theoretically expressed via the parameters of the particle size distributions (PSD). Then, these parameters are determined through least-squares fitting of the line shapes.}
\ifx\pdfoutput\relax\let\pdfoutput=\undefined\fi
\newcount\msipdfoutput
\ifx\pdfoutput\undefined\else
\ifcase\pdfoutput\else
\ifx\paperwidth\undefined\else
\ifdim\paperheight=0pt\relax\else\pdfpageheight\paperheight\fi
\ifdim\paperwidth=0pt\relax\else\pdfpagewidth\paperwidth\fi
\fi\fi\fi
\begin{document}

\title{Nuclear forward scattering in particulate matter: dependence of lineshape on
particle size distribution}
\author{A.I. Rykov\inst{1,2}}
\maketitle

\section{\bigskip1.Introduction\protect\linebreak\protect\linebreak1.1.
Synchrotron M\"{o}ssbauer spectroscopy and lengths d\allowbreak istribution}

Electronic and structural properties of materials were widely studied by
M\"{o}ssbauer spectroscopy in recent 50 years. Nuclear probe placed into deep
structure scale interacts with its environment enabling to quantify the
hyperfine interactions with unprecedented energy resolution. Cutting edge time
domain M\"{o}ssbauer spectroscopy using a synchrotron source has a potential
of even twice better resolution. This is because the synchrotron M\"{o}ssbauer
spectra show no linewidth contribution from radioactive source, and only the
linewidth contribution of absorber sample remains. For this reason the spectra
obtained in time domain and Fourier-transformed to frequency domain might have
narrower linewidths compared to conventional energy spectra measured in
velocity scale.

Another attractive feature of the time-domain M\"{o}ssbauer spectroscopy, to
be explored in this work, consists in the possibility of studying the
particulate matter. The point is that the length of the pathway through the
resonant media is also reflected in nuclear resonant scattering\cite{Kohn1}.
The NFS theory \cite{Kohn1} predicts in time domain the occurrence of the
dynamic beats with periods depending on the lengths of radiation pathway. The
pathway-dependent beats interfere with quantum beats originating from the
quantum hyperfine splitting of nuclear levels.

In this work, the possibility of obtaining the information on the particle
size distribution (PSD) from NFS is explored. Distribution of electronic and
nuclear densities can be characterized by small-angle scattering (SAS) of
x-rays and neutrons, respectively. In contrast, the unique feature of the
method proposed here consists in determining the chemically-specific densities
of \textit{resonant} nuclei. Ferro- or aniferromagnetic ordering sensed by Fe
spins in magnetic materials is probed using the $^{57}$Fe nuclei. Therefore,
our approach is relevant to various natural or synthetic magnetic compounds,
doped by iron isotope, or containing iron intrinsically. Particularly, in
application to multiphase systems, where at least one of the phases contains
iron, there appears a unique chance of getting deeper insight into the
interfacial structure. Thus, the proposed method has the potential to make a
substantial contribution to experimental research both in multi- and
single-phase systems.

Consider the nuclide-embedding particles placed into the synchrotron beam in a
way that excludes the ray shadowing of one particle by another. The averaged
over distribution chord lengths $d(\zeta)$ determines the total intensity of
resonant scattering in forward direction.

In SAS, the distributions of nuclear or electronic densities $d_{\text{n}%
}(\zeta)$, $d_{\text{e}}(\zeta)$ are proportional to the second derivative of
the SAS correlation function $\gamma_{0}(r)$:%

\begin{equation}
d_{\text{n,e}}(r)=\frac{1}{\overline{\zeta}}\frac{d^{2}\gamma_{0}(r)}{dr^{2}}%
\end{equation}
Here $\gamma_{0}(r)=\gamma(r)/\gamma(0)$ and $\gamma(r)$ is self-convolution
of density distribution\cite{Svergun}. The proportionality factor
$1/\overline{\zeta}$ is the inverse mean length among all the particle chords
distributed between $0$ and the maximum particle dimension $D$:%

\begin{equation}
\overline{\zeta}=\int_{0}^{D}\zeta d(\zeta)d\zeta
\end{equation}

In our case, $\overline{\zeta}$ is the average thickness of the resonant
media. The condition of "no shadowing" makes the chord lengths distribution
(CLD) of randomly oriented particles equivalent to the resonant-media
thickness distribution (RTD). In samples without mixing the resonant and
non-resonant media, the RTD is quite analogous to the thickness distribution
introduced previously in x-ray spectroscopy for thickness-related corrections
of x-ray absorption spectra (XAS)\cite{Bausk}. While the thickness
distribution determines the XAS and NFS spectra, the CLD is known to underlie,
except SAS, a variety of properties of porous matter: conductivity, transport
and relaxation. Investigating the CLD is subject matter of a special geometric
sphere of knowledge\cite{Gille}.

\section{1.2. Frequency-domain NFS spectra: the lineshapes}

In this work, we limit ourselves to the case of full coincidence between
thickness and chord lengths distributions (RTD$\equiv$CLD). For example, this
is the case of single layer of particles spread over a surface or
M\"{o}ssbauer particles well diluted in non-M\"{o}ssbauer media to avoid
shadowing of one particle by another. This case includes not only the
spherical particles of various particle size distributions (PSD's), but also a
network of wires or filaments viewed as discs in the plane perpendicular to
the beam propagation direction. The relationships between the PSD, CLD and
curves of SAS are known as solutions of the corresponding inverse scattering
problems\cite{Svergun}. In a similar way, we will establish the relationship
between PSD, CLD and the parameters of lineshape in the Fourier-transformed
NFS spectra, such as shown in Fig. 1 (b).%

\begin{figure}
[ptb]
\begin{center}
\includegraphics[
height=4.5222in,
width=3.3806in
]%
{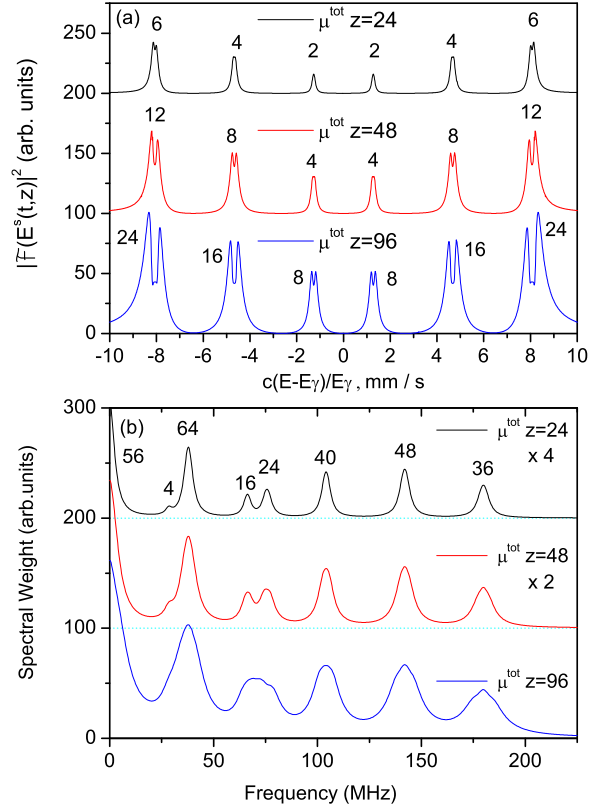}%
\caption{(a) Energy dependence of the NFS intensity in case of scatterer with
the sextet of resonance lines of the transition probabilities 3:2:1:1:2:3. Due
to the multiple scattering each resonance shows the double-hump structure at
large $z$. The spectrum was obtained by squaring the absolute value of the
Fourier-transformed sum of Eq. (3). It was verified that exactly same spectra
can be obtained using the formula (Eq. \ref{Rs}) for squared transmission
amplitude: $\left\vert 1-R^{s}(E)\right\vert ^{2}=\left\vert 1-\exp\left(
-i\sum_{l}\frac{\xi_{l}\Gamma_{0}}{E-E_{l}+i\Gamma_{0}/2}\right)  \right\vert
^{2}.$ Here $\Gamma_{0}$ is the linewidth ( $0.097$ mm/s in $^{57}$Fe). The
individual M\"{o}ssbauer thicknesses $\mu_{l}z$ for each nuclear transition
are indicated. Total dimensionless thicknesses of 24, 48 and 96 are given by
the sum over all transitions $\mu^{\text{tot}}z=\Sigma\mu_{l}z$. (b):
Frequency dependence of the NFS Fourier spectra, obtained using the same
parameters of thickness and hyperfine interactions (magnetic hyperfine field
$H_{\text{hf}}$=500 kOe, quadrupole splitting $\varepsilon=0$).}%
\end{center}
\end{figure}

Our approach to the NFS data treatment in frequency domain is very different
from the conventional one, that has been known since the first observations of
quantum beatings\cite{Gerdau,Hastings,Buerck}. In the frequency domain, we
show that a single quantum beat can be disentangled from the complex sum of
beatings at the condition of large enough hyperfine fields ($H_{\text{hf}%
}\gtrsim200$ kOe). Fitting the lineshape for any individual line of the
Fourier spectra permits to determine the RTD parameters even for complex
distributions of thickness of the resonant media.%

\begin{figure}
[ptb]
\begin{center}
\includegraphics[
height=4.425in,
width=3.3109in
]%
{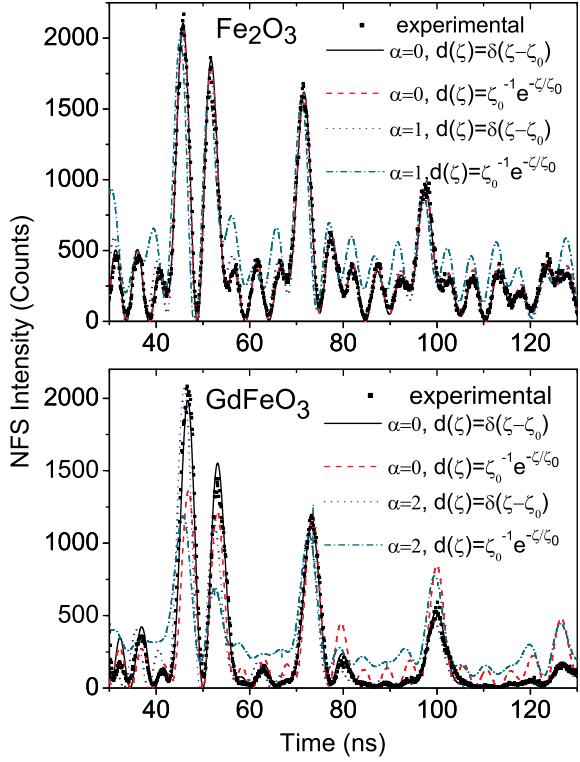}%
\caption{NFS spectra of Fe$_{2}$O$_{3}$ and GdFeO$_{3}$ fitted using the
theoretical intensities derived from the Eqs. (\ref{appr}) and (\ref{QBshift}%
). The goodness of fit is always best for $\alpha=0$. Except the experimental
spectra and fitted theoretical spectra for $\alpha=0$, the simulated spectra
for $\alpha=1$ nm/$\mu$m (Fe$_{2}$O$_{3}$) and $\alpha=2$ nm/$\mu$m
(GdFeO$_{3}$) are also shown. The unenriched oxides Fe$_{2}$O$_{3}$ and
GdFeO$_{3}$ contain the density of the $^{57}$Fe nuclei by 100 and 230 times
smaller than that in bcc metal of $\alpha-^{57}$Fe. }%
\end{center}
\end{figure}
\begin{figure}
[ptb]
\begin{center}
\includegraphics[
height=4.4807in,
width=3.2993in
]%
{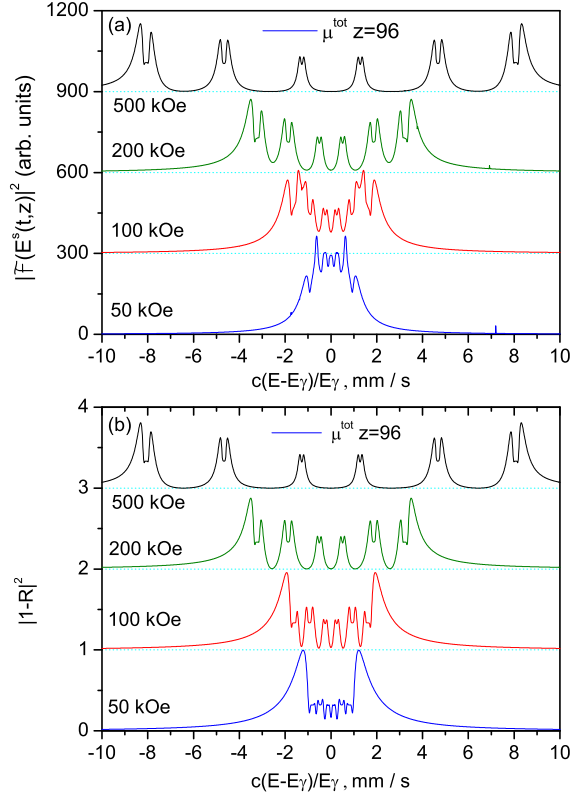}%
\caption{Energy spectra of the scattered radiation corresponding to
approximate (a) and exact (b) solutions of the wave equation for radiation
pulse propagating through resonant media. The squared amplitudes of radiation
field vs. energy are calculated using the Eqs. (\ref{appr}) and (\ref{Rs}) in
(a) and (b), respectively. Note the coincidence of spectra in (a) and (b) for
$H_{\text{hf}}$ $\gtrsim$ 200 kOe.}%
\end{center}
\end{figure}

\section{\protect\linebreak3.Prerequisites\protect\linebreak\protect\linebreak%
3.1. The approximation of singular magnetic hyperfine fields}

Already the first studies of NFS\cite{Shvyd'ko} showed that a phenomenon of
coupling between quantum beats and dynamical beats often takes place at large
lengths of radiation pathways through the resonant media. This hybridization
results from broad magnetic-hyperfine-field distributions that have a strong
effect on the time evolution of the intensity of the forward scattered
radiation. This is the case of magnetic alloys having essentially disordered
atomic structure, such as Invar, or fully amorphous alloys and compounds.
Distribution of quadrupole parameters in nanoparticles or in deformed crystals
is another example of inhomogeneous broadening that gives rise to the coupling
between quantum and dynamical beats. In what follows, we disregard these cases
to focus on the case of singular value of hyperfine magnetic fields for all
the $^{57}$Fe nuclei involved.

\section{2.2. The approximation of large magnetic hyperfine fields}

Even for a singular values of hyperfine fields the radiation scattered forward
in samples of large thickness exhibits a broad energy distribution around each
hyperfine transition, showing the so-called double-hump picture of a hyperfine
transition (Fig.1, a). This picture arises from solving the wave equation for
propagation of a radiation pulse through a resonant medium\cite{Shvyd'ko}.
Closed analytic solutions of corresponding wave equations exist only for a
single-line resonance. However, if we remain in the limit of large energy
separation between the hyperfine transitions, there exists a good
approximation to the analytic solution even in the case of multiple
resonances. This approximation is applicable when the energy separation
between different transitions is large enough compared to natural linewidth
\cite{Smirnov,BuerckHyp,Shvyd'ko}.

Then the approximate solution for the field amplitude of the radiation
scattered in forward direction can be expressed as follows\cite{Shvyd'ko}:
\begin{equation}
E^{s}(t,z)\propto\sum_{l}\mu_{l}z\exp(-\frac{i}{\hslash}\Delta E_{l}%
t-\frac{\tau}{2})\frac{J_{1}(\sqrt{\mu_{l}z\tau})}{\sqrt{\mu_{l}z\tau}}
\label{Ampl}%
\end{equation}

Here the index $l$ numerates the nuclear transitions with the energy
difference of $\Delta E_{l}$ between ground and excited states. The factor
$e^{-\tau/2}$ describes the free-nucleus decay. The dimensionless time
$\tau=t/t_{0}$ is expressed in units of free decay time $t_{0}$ (e.g., $t_{0}%
$=141.1 ns for $^{57}$Fe). The nuclear absorption coefficient $\mu_{l}=$
$\sigma_{0}\rho f_{\text{LM}}\eta w_{l}$ is expressed individually for each
transition using the weight factor $w_{l}$ of the $l$-th transition. The
factor $w_{l}$ takes into account the probability partition between nuclear
sublevels split by hyperfine interactions, so that $\Sigma_{l}w_{l}=1$. Other
factors are the resonance cross section $\sigma_{0}$, the density of the
resonant nuclei $\rho$, the Lamb-M\"{o}ssbauer factor $f_{\text{LM}}$, and the
resonant nuclide isotope abundance $\eta$.

\medskip The time and space variables are entangled in the Eq.(\ref{Ampl}) via
the complex argument of the Bessel function of first kind and order one:
\begin{equation}
J_{1}(\sqrt{\mu_{l}z\tau})=\sum_{k=0}^{\infty}\frac{(-1)^{k}}{k!(k+1)!}\left(
\frac{\sqrt{\mu_{l}z\tau}}{2}\right)  ^{2k+1}%
\end{equation}

Using the variable $\xi_{l}=\mu_{l}z/4$ the power series of the Bessel
argument $\sqrt{\mu_{l}z\tau}/2$ can be replaced by the power series of
$\sqrt{\xi_{l}\tau}$. The dimensionless parameter individual for each $l-$th
transition $\xi_{l}$ is called the transition effective resonant thickness,
resulting in radiation field:%
\begin{equation}
E^{s}(t,z)\propto\sum_{l}\xi_{l}\exp(-\frac{i}{\hslash}\Delta E_{l}%
t-\frac{\tau}{2})\frac{J_{1}(2\sqrt{\xi_{l}\tau})}{\sqrt{\xi_{l}\tau}}
\label{appr}%
\end{equation}

In terms of Eq.(\ref{appr}), the spectra shown in Fig. 1 (a) represent the
squared absolute values of the corresponding Fourier transforms $\left\vert
\mathcal{F}%
(E^{s}(t,z))\right\vert ^{2}$. The values of effective resonant thickness
$\mu_{l}z$ are $l-$specific and proportional to the relative intensity of the
M\"{o}ssbauer line for the $l$-th transition. Three values of total
M\"{o}ssbauer thickness ($\mu^{\text{tot}}z=\Sigma_{l}\mu_{l}z=24,48$ and 96)
are used to draw the Fig.1. The magnetic hyperfine field of 500 kOe and zero
quadrupole splitting are assumed. Very close hyperfine parameters were
observed previously in GdFeO$_{3}$\cite{Rykov}. Spectral weight of the quantum
beat frequency spectra shown in Fig.1 (b) represents the $\cos$-Fourier
transform of the squared sum in right-hand of Eq.(\ref{Ampl}). Information
contained in such Fourier images is the same as in experimental data. In what
follows, we elaborate the concept of lineshape in these Fourier spectra.

\section{2.3. Are the shifts of quantum beat occuring in forward scattering?}

The (Eq.\ref{appr}) was introduced in 1998 by Shvyd'ko et al (see, Eq. (16) in
the work \cite{Shvyd'ko}). In the Eq.(\ref{appr}), the radiation field
$E^{s}(t,z)$ exhibits at the time $t=0$ similar phase for each $l^{\text{th}}$
transition. The same assumption was used in Refs. \cite{Hastings}%
,\cite{Rykov,Smir2,SB}. However, some of the previous works
\cite{Buerck,Smirnov} supposed the existence of a finite difference between
the initial phases (at $t=0$) for different transitions (see the Eq.(29) in
Ref.\cite{Smirnov}):
\begin{equation}
E_{\alpha}^{s}(t,z)\propto\sum_{l}\xi_{l}\exp\left(  -\frac{i}{\hslash}\Delta
E_{l}t-\frac{\tau}{2}+\alpha\sum_{m}\frac{\Gamma\xi_{m}}{\Delta E_{lm}%
}\right)  \psi_{l} \label{QBshift}%
\end{equation}

Here $\psi_{l}$ is the same factor as in Eq.(\ref{appr}), $\psi_{l}$
$=J_{1}(2\sqrt{\xi_{l}\tau})/\sqrt{\xi_{l}\tau}$ and $\Delta E_{lm}$ is the
energy difference between $l^{\text{th}}$ and $m^{\text{th}}$ transitions. In
the experimental data for magnetized $^{57}$Fe foils, the finite quantum beat
shifts were claimed in two early works\cite{Buerck,Smirnov}, but not reported
in precedent\cite{Hastings} and succeeding\cite{Shvyd'ko,Rykov,SB,Smir2}
studies. Quite large values of the parameter $\alpha$ were reported for the
NFS with a single QB component, $\alpha=0.133$ nm/$\mu$m \cite{Buerck} and
$\alpha=0.14$ nm/$\mu$m\cite{Smirnov} If such QB shifts were present for some
of the QB components from sextet, the shape of the curve of sum-total NFS
would distinctly depend on the thickness even in the absence of any
distribution of thickness. In case of thickness distribution, the shift is
expected to smear the quantum beats. On the other hand, no smearing is
expected in case of validity of the zero-shift formula (Eq.\ref{appr}), which
accounts for the thickness distribution only through the envelopes $\psi_{l}$.

No QB shifts were assumed previously at fitting the NFS data for powdered
antiferromagnetic oxides\cite{Rykov}. Let us compare which of the Eqs.
(\ref{appr} and \ref{QBshift}) fits better the experimental time spectra. In
Fig 2, the NFS spectra of Fe$_{2}$O$_{3}$ and GdFeO$_{3}$ are fitted using
both Eqs. (\ref{appr} and \ref{QBshift}). The goodness of fit is always best
for $\alpha=0$. Except the experimental spectra and theoretical spectra for
$\alpha=0$, the spectra for $\alpha=1$ nm/$\mu$m (Fe$_{2}$O$_{3}$) and
$\alpha=2$ nm/$\mu$m (GdFeO$_{3}$) are also shown. Both these oxides were
prepared with natural abundance of $^{57}$Fe. The density of $^{57}$Fe nuclei
in such unenriched Fe$_{2}$O$_{3}$ and GdFeO$_{3}$ oxides is lower than in
$^{57}$Fe foil by the factors of 100 and 230, respectively. For each pair of
transitions, $l$ and $m$, the terms $\alpha\Sigma_{m}(\Gamma\xi_{m}/\Delta
E_{lm})$ are relatively small. In case of magnetic sextet of transitions with
small quadrupole shift, there are three groups of sums $\Sigma_{m}(\Gamma
\xi_{m}/\Delta E_{lm})$ with relative values of $\simeq0$, $\simeq1$, and
$\simeq2$. Each of these terms $\alpha\Sigma_{m}(\Gamma\xi_{m}/\Delta E_{lm})$
is smaller than the QB shift for $\alpha=0.133$ nm/$\mu$m in $^{57}$Fe foil at
least by an order of magnitude. Despite very small values of $\alpha$ assumed,
Fig. 2 shows strong divergence between experimental and simulated for
$\alpha\neq0$ spectra. Shape of the time dependence of the squared field
amplitude $E_{\alpha}^{s}(t,z)$ calculated from Eq.(\ref{QBshift}) exhibits
very strong changes even for small deviations of $\alpha$ from 0. Such a
variation of shape is the best measure for the value of $\alpha$. No
significant deviation of $\alpha$ from $0$ is observed in our data shown in
Fig.2. This is contrasting to the interpretations derived from the data in
magnetized $^{57}$Fe foil with just two unsuppressed transitions\cite{Buerck}.
The complex shape of the NFS time spectra for magnetic sextet is a more
reliable observation\ than the shift of QB reported for the magnetized $^{57}%
$Fe foil doublet. In fact, experimental NFS data suffer generally from the
uncertainty of time origin. This can be caused, for example, by a finite
spread of prompt pulse. This uncertainty may lead to a misleading shift of
spectra as a whole, however, may not change the shape of the NFS pattern.

Thickness distribution modeled by the exponential function $d(\xi)=$ $\xi
_{0}^{-1}\exp(-\xi\xi_{0}^{-1})$ is consistent with the NFS spectra of
Fe$_{2}$O$_{3}$, but inconsistent with the NFS spectra of GdFeO$_{3}$ (Fig.2).
Except difference in quadrupole splitting (cf. $\varepsilon=-0.1$ mm/s in
Fe$_{2}$O$_{3}$ and $\varepsilon=+0.03$ mm/s in GdFeO$_{3}$ \cite{Rykov})
these samples differ by factor 4 in their M\"{o}ssbauer thickness and by
factor of 10 in their metric thickness. The thicker sample GdFeO$_{3}$ showed
a disagreement with the assumption of exponential distribution. In such a
thick sample, the particles were most probably stacked in several layers,
shadowing each other. Resulting distribution is much closer to $\delta
$-function than to exponential distribution. We obtained $\xi_{0}^{{}}%
\simeq22$ assuming the uniform thickness, $d(\xi)=\delta(\xi-\xi_{0}^{{}})$.
By contrast, in the thinner sample ( Fe$_{2}$O$_{3}$), the distribution of
thicknesses about the mean value $\xi_{0}^{{}}$was relatively broad. As the
result, the fitting quality was equally good for $d(\xi)=\delta(\xi-\xi
_{0}^{{}})$ and $d(\xi)=$ $\xi_{0}^{-1}\exp(-\xi/\xi_{0}^{{}})$ when the
fitting is done in the time range between 30 and 130 ns (Fig .2). However, the
exponential distribution suits better when the spectra were fitted in broader
range. The parameters can be fitted either in time domain or in
Fourier-transformed spectra. Below we present fitting the Fourier spectra as
the most convenient and transparent method.

\section{2.4. Exact energy spectra of the scattered radiation compared with
the large-fields approximation}

The condition of validity of the approximation (Eq.\ref{appr}) was given in
1998 by Shvyd'ko et al\cite{Shvyd'ko}. It was pointed out\cite{Shvyd'ko} that
the Eq.(\ref{appr}) is valid even in the thick samples, if the nuclear
transition energies $E_{l}$ are well separated from each other. This means
that the condition of validity of Eq.(3) is the greatness of the magnetic
hyperfine field compared to natural Lorentzian linewidth. Numerically, this
statement is verified and confirmed in Fig.3. Our calculations aimed to answer
the question: how large fields are sufficient to apply the approximation? In
Fig.3, the squared value of $\left\vert 1-R^{s}(E)\right\vert $ is
plotted,\ where $R^{s}(E)$ being the system response function:%
\begin{equation}
R^{s}(t,z)=\exp\left(  -i\sum_{l}\frac{\xi_{l}\Gamma_{0}}{E-E_{l}+i\Gamma
_{0}/2}\right)  \label{Rs}%
\end{equation}
The Eq.(\ref{Rs}) is the exact result, therefore, the Fig.3 presents the
comparison between the approximate result $\left\vert
\mathcal{F}%
(E^{s}(t,z))\right\vert ^{2}$ (a) and the exact one $\left\vert 1-R^{s}%
(E)\right\vert ^{2}$ (b). Large value of thickness $\mu^{\text{tot}}z=96$ was
chosen to plot the Fig.3. Full coincidence of the exact result and analytic
solutions is shown for $H_{\text{hf}}\gtrsim200$ kOe.

\bigskip

\section{\protect\linebreak3.Results\protect\linebreak\protect\linebreak3.1.
Lineshapes of individual lines of the Fourier spectra for hyperfine sextet}

In large hyperfine fields ($H_{\text{hf}}\gtrsim200$ kOe), the difference
between numeric and analytical solutions is negligible. Therefore, the squared
sum over hyperfine transitions (Eq.\ref{Ampl}) can be converted into sum over
the intertransitional pairs:
\begin{equation}
I(t,z)\propto z^{2}e^{-\tau}\sum_{n}A_{n}(z,\tau)\cos\varpi_{n}t \label{NFS}%
\end{equation}
Fourier transformation of the Eq.(\ref{NFS}) results in the quantum beat
frequency spectra shown in Fig.1 (b).\ In case of pure magnetic hyperfine
interactions, the quantum beats frequencies $\varpi_{n}$ and amplitudes
$A_{n}(z,\tau)$ can be expressed via the parameters $\Omega=%
\frac14
g_{3/2}\mu_{N}H_{\text{hf}}/\hbar$ and $\psi_{j}=J_{1}(2\sqrt{j\xi_{\min}\tau
})/\sqrt{j\xi_{\min}\tau}$ $(j=1,2,3)$ , respectively, as shown in the Table 1.

\bigskip\medskip

Table 1. The quantum beat frequencies $\varpi_{n}$ and amplitudes
$A_{n}(z,\tau)$ for nuclear levels split by purely magnetic hyperfine
interactions. The frequencies are expressed through $\Omega=%
\frac14
g_{3/2}\mu_{N}H_{\text{hf}}/\hbar$ and the amplitudes are expressed through
$\psi_{j}=J_{1}(\sqrt{j\mu z\tau})/\sqrt{j\mu z\tau}$ $(j=1,2,3).\smallskip$

\begin{center}%
\begin{tabular}
[c]{|ll|}\hline
$\varpi_{n}$ & \multicolumn{1}{|l|}{$A_{n}(z,\tau)$}\\\hline
0 & $\psi_{1}^{2}+4\psi_{2}^{2}+9\psi_{3}^{2}$\\
3$\Omega$ & $\psi_{1}^{2}$\\
4$\Omega$ & $4\psi_{1}\psi_{2}+12\psi_{2}\psi_{3}$\\
7$\Omega$ & $4\psi_{1}\psi_{2}$\\
8$\Omega$ & $6\psi_{1}\psi_{3}$\\
11$\Omega$ & $6\psi_{1}\psi_{3}+4\psi_{2}^{2}$\\
15$\Omega$ & $12\psi_{2}\psi_{3}$\\
19$\Omega$ & $9\psi_{3}^{2}$\\\hline
\end{tabular}
\medskip

\medskip
\end{center}

There are eight spectral lines corresponding to eight nonidentical
intertransitional energy differences for purely magnetic hyperfine
interactions. According to the Table 1, each line in the frequency domain has
its own lineshape. The line at highest frequency is the widest one. This line
turns out to be most suitable for our analysis because the Fourier
transformation of $\psi_{3}^{2}$ $=J_{1}^{2}(2\sqrt{\xi\tau})/\sqrt{\xi\tau
}=J_{1}^{2}(2\sqrt{3\xi_{\min}\tau})/3\xi_{\min}\tau$ can be found
analytically. Here we denoted by $\xi$ and $\mu z$ the transition-specific
quantities $\xi=$ $3\xi_{\min}=\mu z/4=\mu^{\text{tot}}z/16.$By the minimum
thickness, $\xi_{\min}=\mu^{\text{tot}}z/48$ we denoted the thickness of the
weakest transition (1 in the notation 3:2:1:1:2:3).

In presence of quadrupole interactions some of these eight lines split into
doublets and triplets to produce fourteen lines, however, the line of highest
frequency always stands alone. This outstanding feature makes it easy to apply
the suggested here analysis in frequency domain. It is from this line the
parameters of thickness distribution are derivable most easily.

\section{3.2. The shape and width of highest-frequency line}%

\begin{figure}
[ptb]
\begin{center}
\includegraphics[
height=2.6899in,
width=3.3698in
]%
{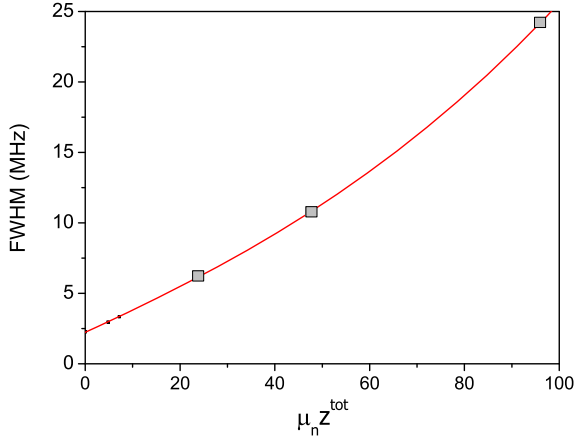}%
\caption{Thickness dependence of the full-width-at-half-maximum (FWHM) of the
highest-frequency line shown in Fig.1(b). The lineshape is the convolution of
the thickness-independent Lorentzian and thickness-dependent term $\left\vert
1-_{2}\digamma_{3}[\{\frac{1}{3},\frac{3}{4}\},\{\frac{1}{2},1,\frac{3}%
{2}\},-\frac{\mu^{2}z^{2}}{4\omega^{2}}]\right\vert $ with $_{2}\digamma_{3}$
being the hypergeometric function. Pure Lorentzian contribution to the
linewidth at $z=0$ is 2.26 MHz. }%
\end{center}
\end{figure}
\begin{figure}
[ptb]
\begin{center}
\includegraphics[
height=2.5098in,
width=3.3698in
]%
{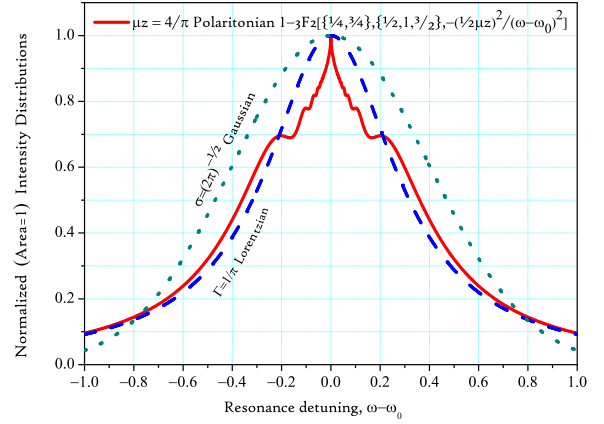}%
\caption{Comparison of the universal distribution functions normalized by both
area and height: the "Polaritonian", obtained as Fourier transform of the QB
envelope $\mathcal{F}(\psi_{3}^{2})=$ $1-_{3}F_{2}[\{\frac{1}{4},\frac{3}%
{4}\},\{\frac{1}{2},1,\frac{3}{2}\},-\frac{\mu^{2}z^{2}}{4(\omega-\omega
_{0})^{2}}]$(continuous line), compared to Lorentzian $\frac{1}{\pi\Gamma
}\frac{1}{1+(\omega-\omega_{0})^{2}/\Gamma^{2}}$(dashed line) and Gaussian
$\frac{1}{\sqrt{2\pi\sigma^{2}}}\exp(-\frac{(\omega-\omega_{0})^{2}}%
{2\sigma^{2}})$ (dotted line). }%
\end{center}
\end{figure}

In Eq. (\ref{NFS}), except the functions $\psi_{j}$, the factor $e^{-\tau}$
contributes the lineshapes. Unlike the $\psi_{j}-$dependent factors individual
for each line, the factor $e^{-\tau}$ is common for all NFS data, therefore,
it can be easily removed by the sharpening procedure as shown below. Here we
note that the $e^{-\tau}$-factor dominates the full-width-at-half maximum for
small linewidths (FWHM $\lessapprox$ 5 MHz). Remaining factors shown in the
Table 1 become predominant for larger linewidths. The Fourier images of these
remaining factors are the $\Lambda-$shaped functions with some shoulders. In
Fig.1(b), the shoulders start to be visible for the singlet line as the
thickness increases. In the limit of zero thickness the Lorentzian
contribution to the linewidth is 2.2575 MHz (Fig. 4). The plot of FWHM vs
thickness is not linear because the thickness-dependent component differs in
shape from the Lorentzian $L=%
\mathcal{F}%
(e^{-\tau})$.\medskip

Since the NFS spectra are defined for $t>0$ \ the Fourier-transform is defined
as $%
\mathcal{F}%
(f)=\sqrt{2/\pi}\int_{0}^{\infty}f(t)\cos(\omega t)dt$. For the sake of
normalization of the line intensities in Fourier spectra we introduce the new
variable $\zeta$ that is another useful parameter related to M\"{o}ssbauer
thickness:
\begin{equation}
\zeta=\pi\xi=\frac{\pi}{4}\mu z=\frac{\zeta^{\text{tot}}}{4}=\frac{\pi}{16}%
\mu^{\text{tot}}z\text{ .} \label{ch}%
\end{equation}
The new parameter $\zeta$ presents the measure of FWHM in the NFS Fourier
spectra vs. angular frequency $\omega$. In contrast, the parameter $\xi$ is
the suitable measure of FWHM in the NFS Fourier spectra vs. linear frequency.
Transforming the NFS time spectra, Eq. (\ref{NFS}), to the frequency domain
one obtains a set of lines, whose lineshapes are represented by the
convolution of the Fourier images of the free decay component and the
thickness-dependent component. First component is the free-decay Lorentzian
$L=%
\mathcal{F}%
\lbrack\exp(-\tau)]$ independent of the polaritonic parameter $\zeta$. Second
is the polaritonic component related to propagation of radiation pulse through
the resonant media. Because of this convolution the full lineshape can be
found only by numerical calculations. For a single-line spectrum this was
already done\cite{Rykov}. The spectral lineshape was shown to evolve with
increasing thickness from the Lorentzian profile to a different profile
(so-called $\Lambda$-shaped \cite{Rykov}). However, no analytic expression was
obtained yet for the thickness-dependent component of the lineshape. This
solution will be found in the present work. The method of determination of the
thickness distribution parameters simply follows from this analytic solution.

In very thin samples, the same analysis could be applicable to the line at
$\varpi_{n}=3\Omega\ $shaped as simply as the Fourier-image of $\psi_{1}^{2}$,
however, this line is just a tiny satellite of the strongest line of the NFS
Fourier spectra (see Fig.1,b). That is why the operational range of
thicknesses for this line is narrower than that for the line at $\varpi
_{n}=19\Omega$. In sharpened NFS Fourier spectra, we could manage to find the
analytic solution for the spectral shape of these two lines, the narrowest one
and the widest one, but not for the other lines.%

\begin{figure}
[ptb]
\begin{center}
\includegraphics[
height=2.4906in,
width=3.3532in
]%
{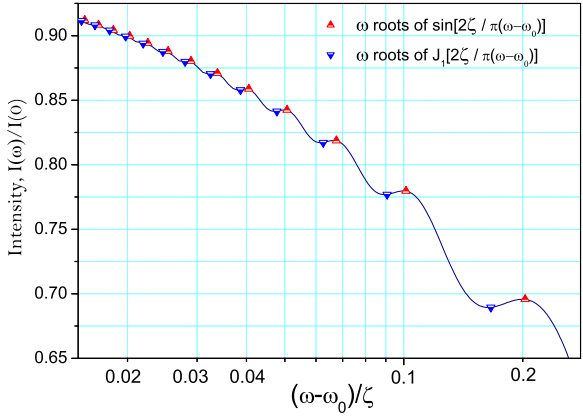}%
\caption{Center of the Polaritonian $1-h_{\frac{3}{2}}\left[  -\frac
{4\zeta^{2}}{\pi^{2}(\omega-\omega_{0})^{2}}\right]  $ vs. $(\omega-\omega
_{0})/\zeta$ (Eq. \ref{FT}) in semilogarithmic scale. Maxima and minima in the
oscillating behavior of the lineshape are indicated by up and down triangles,
respectively.}%
\end{center}
\end{figure}
%

\begin{figure}
[ptb]
\begin{center}
\includegraphics[
height=2.4093in,
width=3.3898in
]%
{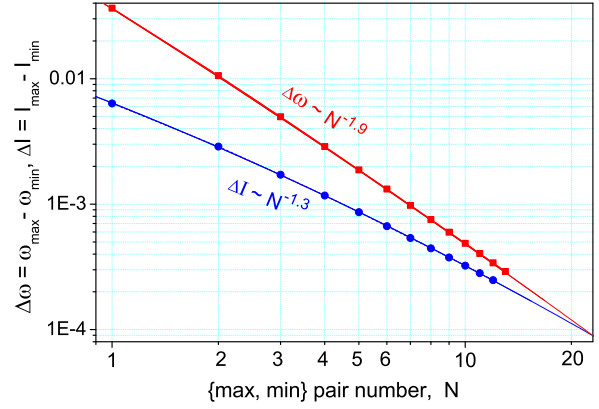}%
\caption{The widths $\Delta\omega$ and the heights $\Delta I$ of the satellite
shoulders in the polaritonic lineshapes shown in log-log scale. The numeration
counter of shoulders $N$ goes from outwards to the center of the spectral
line. Shown on the plot are the power exponents calculated using $N=13$.}%
\end{center}
\end{figure}

\section{3.3 Shape of the polaritonic component for a uniform-thickness layer
(foil)}

Thus, since the sharpened lineshape is well-defined theoretically one can fit
the sharpened data with a thickness-dependent analytic expression. Prior the
Fourier transformation the experimental spectra are multiplied by the factor
$\exp(\tau)$. Then the NFS sum for a uniform thickness $\zeta$ (Eq.\ref{NFS})
will have the highest-frequency term as follows:%
\begin{equation}
\widetilde{I}(t,\zeta)\exp(\tau)\propto\zeta^{2}\left(  \frac{J_{1}\left(
2\sqrt{\zeta\tau/\pi}\right)  }{\sqrt{\zeta\tau/\pi}}\right)  ^{2}\cos19\Omega
t \label{rh}%
\end{equation}
This procedure allowed us to obtain in the frequency domain the lineshape
deconvoluted from the natural decay Lorentzian. Therefore, in the frequency
domain, instead of the intensities $I(\omega),$ calculated
previously\cite{Rykov}, we operate now with the sharpened intensities
$I_{\text{sh}}$ $(\omega,\zeta)=$ $\allowbreak%
\mathcal{F}%
\lbrack\widetilde{I}(t,\zeta)\exp(\tau)]$. The Fourier transform of the
remaining lineshape factor in the right hand side of the Eq.(\ref{rh}) has the
functional form:%

\begin{equation}
\allowbreak%
\mathcal{F}%
\left[  \frac{\zeta}{\sqrt{2\pi}}\left(  \frac{J_{1}\left(  2\sqrt{\zeta
\tau/\pi}\right)  }{\sqrt{\zeta\tau/\pi}}\right)  ^{2}\right]  =1-h_{\frac
{3}{2}}\left(  -\frac{4\zeta^{2}}{\pi^{2}\omega^{2}}\right)  \label{FT}%
\end{equation}
Here $\omega$ is the dimensionless angular frequency corresponding to the
dimensionless time $\tau$. By $h_{\frac{3}{2}}(x)$ we denoted the
hypergeometric function $_{2}\digamma_{3}[\{\frac{1}{3},\frac{3}{4}%
\},\{\frac{1}{2},1,\frac{3}{2}\},x]$, using the last but one argument of the
function as subscript. The value $\zeta=1$ in Eq.(\ref{FT}) correspond to the
normalized, both in area and height, distribution density (Fig. 5). Similarly
normalized Lorentzian and Gaussian distributions are $\left[  1+\pi^{2}%
(\omega-\omega_{0})^{2}\right]  ^{-1}$ and $\exp\left[  -\pi(\omega-\omega
_{0})^{2}\right]  $. The FWHM of the polaritonic function is slightly larger
than FWHM of Lorentzian but smaller than that for Gaussian. To within this
difference caused by the difference of the lineshape, the natural linewidth
caused by the natural decay $\exp(-\tau)$ is equivalent to the thickness of
$\zeta_{\text{eqv=}\tau}=\pi$ or $\xi_{\text{eqv=}\tau}=1$. \ The origin of
$\omega$ is located at the highest quantum beat frequency, hence, the positive
and negative values of $\omega$ are measured relative to $\omega_{0}%
=19\Omega,$ that is not shown in Eq.(\ref{FT}). The same notation for $\omega$
will be used below ($\omega$ "measuring" relative to $\omega_{0}$).

The polaritonic lineshape has a very sharp tip in the center and the wings
decaying similarly to the Lorentzian. It has the main shoulders $I_{1}%
\approx0.7I_{0}$ at $\omega_{1}\approx\omega_{0}\pm0.2\zeta$. A sequence of
less-marked shoulders exists in closer proximity to the resonance. These
shoulders are formed by shallow minima and maxima at $\omega_{N}^{\min}$ and
$\omega_{N}^{\max}$, respectively, as shown in semilogarithmic scale in Fig.
6. The maxima are located at $\omega_{N}^{\max}=\omega_{0}\pm2\zeta/N\pi^{2}$,
$N=1,2..\infty.$The minima are given by the roots of the equation:
\begin{equation}
J_{1}\left(  \frac{2\zeta/\pi}{(\omega_{0}\pm\omega_{N}^{\min})}\right)  =0
\end{equation}

The width of the shoulders $\Delta\omega_{N}=$ $\omega_{N}^{\max}-\omega
_{N}^{\min}$and the heights of the satellite peaks $\Delta I_{N}=$
$I_{N}^{\max}-I_{N}^{\min}$ decrease with N approximately as the power laws
$N^{-1.9}$ and $N^{-1.3}$, respectively. Both curves $\Delta\omega_{N}$ vs.
$N$ and\ $\Delta I_{N}$ vs. $N$ show a small curvature in the log-log scale
(Fig. 7). They merge each other both in position and in slope as $N$ increases
and the oscillations of $I(\omega)$ fade out with $\omega$ approaching
$\omega_{0}$.

The NFS radiation field and the nuclear currents are the polaritonic
subsystems feeding each other as the excitation pulse propagates through the
resonant media\cite{Sm2007,Kohn2}. As the time elapses after the arrival of
prompt pulse of synchrotron radiation the frequency of the energy exchange
between these two subsystems at a fixed coordinate $z$ decreases. The nodes
and antinodes change each other less and less frequently. The region of small
frequencies in the NFS Fourier spectra is dominated by the harmonics from
large $\tau$ of the NFS time spectra. Lorentzian shape in frequency domain
correspond to $\exp(-\tau)$ in time domain and, vice versa, the Lorentzian's
decay asymptotic $\tau^{-2}$ would produce around $\omega_{0}$ the
triangularly peaked shape $\exp(-\left\vert \omega-\omega_{0}\right\vert )$.
In Fig. 5, we observe that the peak is ever sharper than triangular. Taking
the function $\psi_{3}^{2}(\tau)$ at the points of antinode maxima we observe
that $\psi_{3}^{2}(\tau_{\text{a}})$ decays with $\tau$ as $\tau^{-3/2}$,
i.e., even slower than the Lorentzian's asymptotic $\tau^{-2}$. Indeed, one
can check that the Fourier transformation of the function $(1+|\tau|^{\frac
{3}{2}})^{-1}$ gives a very sharp peak, similar to that of Fig. 5, but free of
the oscillations investigated in Figs 6 and 7.

\smallskip

\section{3.4 Deconvolving the polaritonic component using the technique of
sharpening of the experimental Fourier spectra}

Sharpening the NFS Fourier spectra is shown in Fig. 8 for the data collected
from unenriched Fe$_{2}$O$_{3}$ with natural abundance of $^{57}$Fe. The
spectra were first fitted in the experimental time domain $0.18<\tau<3$ and
then extrapolated to the short-delay region, $\tau<0.18$. Missing original
data for the $\tau<0.18$ caused by the prompt flash overshooting detector were
thus restored. The expression for fit function was given previously (see
Eq.(28) in Ref. \cite{Rykov}). Instead of 8 lines for $\varepsilon=0$ there
are actually 14 lines for the combined magnetic and quadrupole Hamiltonian.
The fitted quantum beats parameters are $H_{\text{hf}}=517.5$ kOe and
$\varepsilon=-0.1$ mm/s. Due to the smallness of $\varepsilon$ the lines are
grouped into 2 symmetric triplets, 2 symmetric doublets and four singlets. The
doublets and triplets are clearly distinguished in the sharpened spectra.%

\begin{figure}
[ptb]
\begin{center}
\includegraphics[
height=5.6737in,
width=3.2603in
]%
{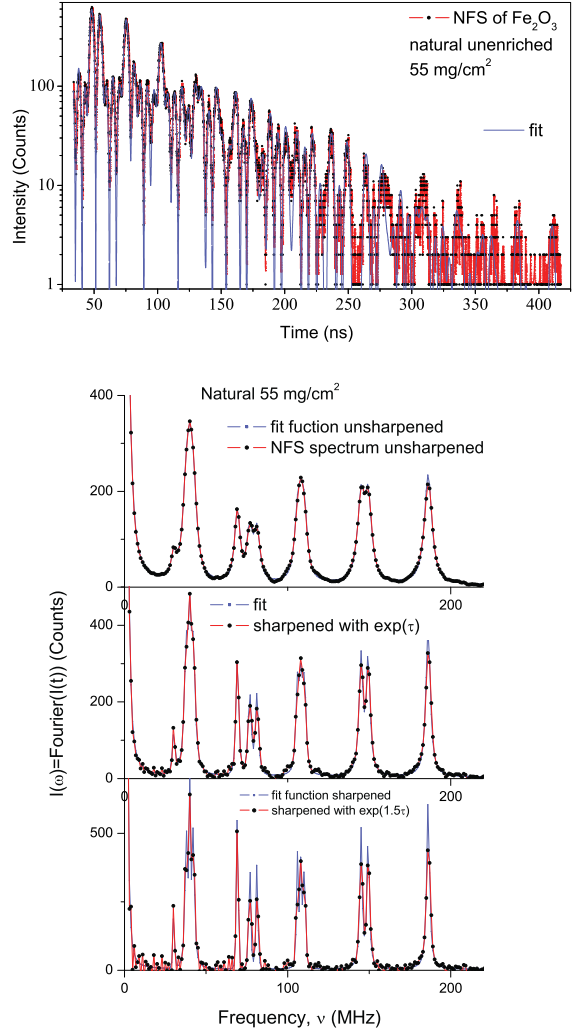}%
\caption{NFS spectra of the Fe$_{2}$O$_{3}$ powder and their unsharpened and
sharpened Fourrier transforms. Top panel data are from \cite{Rykov}. The data
are plotted vs. linear frequency, to be distinguished from $\omega$ in Figs.
1-3. The datapoints in lower panel are obtained via fast Fourier transforms of
NFS time spectra according to formula $I_{j}=\frac{1}{n}\Sigma_{k}%
\widetilde{I}_{k}e^{-i(2\pi j/m)k}$ with the number of channels in NFS spectra
$m$=4096. Solid (blue) line is the sum of Lorentzians of different widths
$\zeta_{\text{eff}}^{2}$ fitted individually for each of 14 lines. Each
Lorentzian is the Fourier image of the time-domain exponential function
fitting the individual quantum beat envelope: $\mathcal{F}\left[  \frac
{\zeta_{\text{eff}}}{\sqrt{2\pi}}\exp(-\zeta_{\text{eff}}\tau/\pi)\right]
=\frac{\zeta_{\text{eff}}^{2}}{\zeta_{\text{eff}}^{2}+\pi^{2}\omega^{2}}.$The
constraints between the parameters of $\zeta_{\text{eff}}$ for different lines
was used to reduce the number of parameters from 14 to one thickness parameter
(see Eq.28 in Ref. \cite{Rykov}). }%
\end{center}
\end{figure}

From the viewpoint of theory the $\exp(-\tau)$-sharpened spectrum in the
middle panel of Fig. 8 must possess the purely polaritonic lineshape. The
factor $\exp(\tau)$ permits us to isolate the polaritonic lineshape in its
undistorted form predicted by the theory. In the middle panel of Fig. 8, the
largest-frequency line at $\varpi_{n}=19\Omega$ possess the largest
polaritonic linewidth $\zeta=3\zeta^{\min}=$ $4.86$.

The sharpening technique is also very useful for the illustration of the
relationship between the width of different spectral lines. To the first
approximation we can neglect the non-linearity in the plot of FWHM vs
thickness in Fig. 4 and assume the additivity of the Lorentzian and
polaritonic linewidths. Therefore, deliberately, the spectrum might be further
sharpened using the exponent larger than $\exp(\tau)$ in Eq.(\ref{rh}). Such a
heuristic sharpening is shown for the sharpening factor $\exp(3\tau/2)$ in the
bottom panel of Fig. 8. The residual linewidth of the narrowest line at
$\varpi_{n}=3\Omega$ \ became very small ($\zeta_{\text{res}}\simeq0.05$).
Such an oversharpening is equivalent to subtraction of $3\pi/2$ from the
linewidth, while the theoretic $\exp(\tau)$-sharpening correspond to
subtraction of $\pi.$ According to the Table 1, the polaritonic linewidths
determined by the function $\psi_{3}^{2}$ is three times larger than the
polaritonic linewidths determined by the function $\psi_{1}^{2}$. This
relationship gives us the following equation for the polaritonic linewidths of
the lines at $\varpi_{n}=3\Omega$ and $19\Omega$:%
\begin{equation}
3\left(  0.05+\frac{\pi}{2}\right)  =4.86
\end{equation}
This relationship is in full agreement with our observation in Fig. 8.

In the middle panel of Fig. 8, there appear only a few experimental points per
FWHM of a spectral line. Two questions are remaining in this respect. First of
all, it is natural to ask how the appearance of spectra can be improved. What
kind of experimental limitations should be lifted over to raise the number of
experimental points per FWHM? Second, we must indicate the way how to apply
the analysis in practice for determination of the particle distribution characteristics.

\section{3.5 Effects of temporal range cut-off and resolution in frequency}

Generally, the interval $\Delta\omega$ between points of the discrete Fourier
spectra is inverse total sample time:%

\begin{equation}
\Delta\omega=\frac{2\pi}{m\Delta\tau} \label{Ince}%
\end{equation}
In NFS, the value $\tau_{\max}=m\Delta\tau$ is limited by the interval between
bunches of electrons of the storage ring. The data of Fig. 8 were collected in
single-bunch regime\cite{Rykov}, with the interval between bunches of 500 ns
and $\tau_{\max}=3.54.$ This experimental restriction on $\tau_{\max}%
=m\Delta\tau$ determines the number of points per FWHM. The number of channels
$m=4096$ defines the NFS sampling frequency. According to the `uncertainty
principle', Eq.(\ref{Ince}), the increase in the value of $m$ will not
influence the number of points per FWHM. However, in very thick samples,
approximately for $\zeta>700$, the existing number of channels $m=4096$ would
provide insufficient sampling frequency for the NFS signal. To obtain the full
set of frequencies in the Fourier spectra, Nyquist-Shannon sampling theorem
prescribes sampling the signal with the frequency twice larger than the
highest frequency of the signal. In this case only, one should increase the
number of channels compared to our value of $4096$.\medskip\medskip

Thus, in Fourier spectra, total sampling time $\tau_{\max}$ limits the number
of points per linewidth, which is proportional to $\zeta$. With increasing
$\zeta$, the broader is a line of Fourier spectra, the narrower is the
function $\psi_{3}^{2}(\tau)=$ $\left(  J_{1}\left(  2\sqrt{\zeta\tau/\pi
}\right)  /\sqrt{\zeta\tau/\pi}\right)  ^{2}$ in time domain. In the
experimental interval $0<\tau<\tau_{\max}$, the number of nodes and antinodes
of the $\psi_{3}^{2}(\tau)$-function increases with increasing $\zeta$. Table
2 shows what should be the value of thickness $\zeta$ to match the
experimental window $\tau_{\max}=3.54$ with the $n$-th node or $n$-th antinode
of the function $\left(  J_{1}\left(  2\sqrt{\zeta\tau/\pi}\right)
/\sqrt{\zeta\tau/\pi}\right)  ^{2}$.

\medskip\smallskip

Table 2. Values of thickness $\zeta$ and $\exp(\tau)-$weighted intensity
$\pi\left(  J_{1}\left(  2\sqrt{\zeta\tau/\pi}\right)  \right)  ^{2}/\zeta
\tau$ for the NFS node and antinode locations at the upper boundary of the
experimental time window $t=500$ ns ($\tau\approx3.54)\ $in a layer of uniform thickness.%

\begin{tabular}
[c]{|llll|}\hline
No. &
\begin{tabular}
[c]{l}%
Node $\zeta$\\
$t=$0.5$\mu$s
\end{tabular}
&
\begin{tabular}
[c]{l}%
Antinode $\zeta$\\
$t=$0.5 $\mu$s
\end{tabular}
&
\begin{tabular}
[c]{l}%
Antinode\\
$\frac{I(t=0.5\mu\text{s})}{I(t=0)}$%
\end{tabular}
\\
1 & 3.26 & 5.85 & 0.0175\\
2 & 10.9 & 15.7 & 4.16$\cdot10^{-3}$\\
3 & 23.0 & 30.0 & 1.60$\cdot10^{-3}$\\
4 & 39.4 & 48.6 & 0.78$\cdot10^{-3}$\\
5 & 60.2 & 71.6 & 0.44$\cdot10^{-3}$\\\hline
\end{tabular}
\medskip\smallskip\smallskip

\medskip In Fig. 9, the function $\left(  J_{1}\left(  2\sqrt{\zeta\tau/\pi
}\right)  /\sqrt{\zeta\tau/\pi}\right)  ^{2}$ is Fourier-transformed to the
frequency domain for the $\zeta$ values from Table 2 after the data cutoff at
$\tau_{\max}=3.54$. Since these Fourier-images are plotted against
$(\omega-\omega_{0})/\zeta$ they are all close to collapsing to the single
master curve (cutoff-free), shown above in Fig. 4.

Three useful types of behavior in the cutoff-affected data (Fig. 9) are of
interest. First, the number of points per FWHM is inverse of $\zeta$ as
discussed above. Second, the discrepancy between the master curve and the
cutoff-affected data culminates in the line center, and decreases towards the
line wings showing an oscillating behavior. Third, as shown in the inset, the
oscillation drops with increasing $(\omega-\omega_{0})$ more rapidly when the
value of $\zeta$ from Table 2 is at node than at antinode of the $\psi_{3}%
^{2}(\tau)$-function. All these simulations were done assuming the homogeneous
thickness. When the resonant media consist of particles the cutoff effects
will be smaller. This is not counterintuitive because the thickness
inhomogeneity would smear the oscillations of $\psi_{3}^{2}(\tau)$. In the
frequency domain, the inhomogeneity of thickness would produce the effects
similar to cutoff, namely, the more smeared oscillations of $\psi_{3}^{2}%
(\tau)$ produce the less sharp vertice in related spectral line. In what
follows, this effect will be compared for several model thickness
distributions. \ %

\begin{figure}
[ptb]
\begin{center}
\includegraphics[
height=2.5031in,
width=3.2395in
]%
{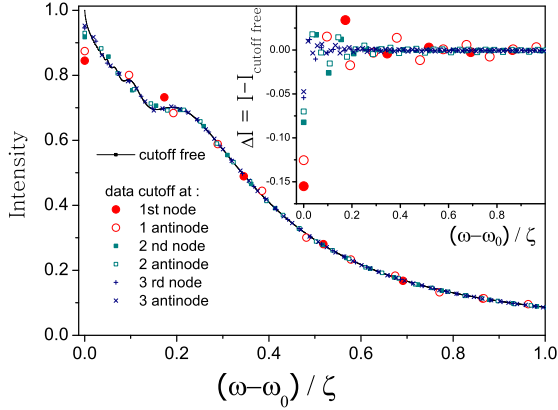}%
\caption{Lineshape profiles illustrating the influence of finiteness of the
experimental time window (cutoff effects) on the lineshape in the NFS Fourier
spectra. The full curve is the cutoff-free data similar that shown in Fig.3.
Points correspond to the Fourier images of the function $\left(  J_{1}\left(
2\sqrt{\zeta\tau/\pi}\right)  /\sqrt{\zeta\tau/\pi}\right)  ^{2}$ cutted off
at $\tau=3.54$ for the values of $\zeta$ from 3 upper lines of Table 2. }%
\end{center}
\end{figure}

\bigskip

\section{3.6 Fitting the thickness as linewidth in sharpened Fourier spectra}

Although our sample was made of particulate Fe$_{2}$O$_{3}$, first we consider
it forming a uniform layer. Within the small-thickness approximation each of
the $\psi_{j}$ function can be expanded into power series of thickness (see
Eq. 28 in Ref. \cite{Rykov}), that simplifies fitting $\xi$ directly in time
domain. The refinement of thickness parameter in this way has resulted in
$\zeta^{\text{tot}}=12\zeta^{\min}=19.5$ ($\xi^{\text{tot}}=6.19$). This total
thickness is slightly larger than the value of $\xi_{\text{w}}^{\text{tot}%
}=4.13$ calculated from the sample weight of 55 mg/cm$^{2}$. The difference
results from the inhomogeneity of the thickness. This inhomogeneity is taken
into account below using two approximations: (i) large spherical micronic-size
particles; (ii) smaller particles of Fe$_{2}$O$_{3}$ (nanoparticles) filling
the space between the larger particles of non-resonant media (MgO filler).%

\begin{figure}
[ptb]
\begin{center}
\includegraphics[
height=6.6334in,
width=3.1731in
]%
{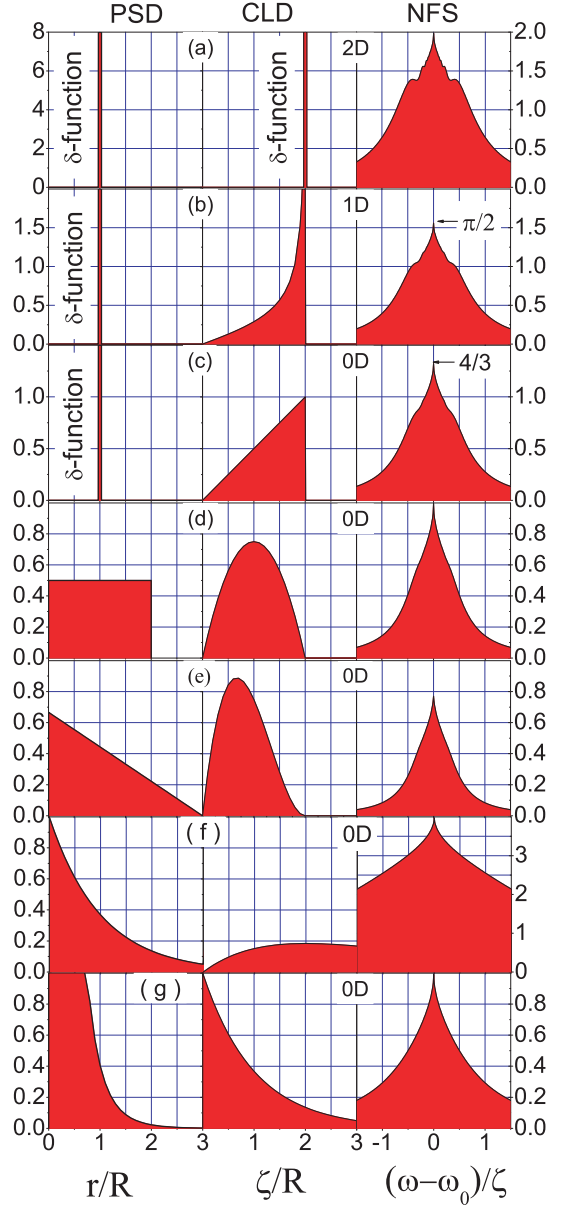}%
\caption{Size distributions (left column), thickness distributions (middle),
and NFS lineshapes (right) for 2D layer (a), 1D wires (b), uniformly-sized
spheres (c), spheres with rectangular (d), triangular (e), exponential (f)
size distributions and with the size distribution of the form $F(r)=(R/r^{2}%
+2/r)\exp(-2r/R)$ (g). In (a) $R$ is the layer half-thickness, elsewhere $R$
is the average radius of sphere or wire.}%
\end{center}
\end{figure}

\bigskip

\section{3.7. Thickness distribution and lineshape for filaments or wires of
the radius $R$}

In the particulate matter, owing to the distribution of the chord lengths the
theoretical NFS lineshape acquires the forms shown in the last columns of Fig.
10. In first and second columns, the PSD and CLD are shown, respectively. The
hypergeometric function of Eq.(\ref{FT}) is reproduced in (a) in the same raw
with the uniform thickness layer PSD and CLD, both represented by $\delta
$-functions. The second raw (b) correspond to the filaments or wires of
uniform radius. They are placed into the plane perpendicular to the beam
direction, therefore, their CLD correspond to the distribution of the chord
lengths $\zeta$ of the discs in 2D. For a disc of the size $2R$ using the
running impact parameter $b$ we can express $\zeta=2\sqrt{R^{2}-b^{2}}$and
$b=\sqrt{R^{2}-\zeta^{2}/4}$. Similarly to thickness distribution \cite{Bausk}
the CLD $d(\zeta)$ can be found from the relationship $d(\zeta)=dS/S$. Here
$dS$ is the relative part of the sample having thickness $\zeta.$ Therefore,
the disc CLD is $d(\zeta)=$ $S^{-1}(dS/d\zeta)=b^{-1}\left\vert db/d\zeta
\right\vert =\zeta(4R)^{-1}(R^{2}-\zeta^{2}/4)^{-%
\frac12
}$. The lineshape can be obtained through integration of the disc CLD with the
right-hand side of the Eq. (\ref{FT}) and $\zeta$ from the Eq. (\ref{rh}):%

\begin{equation}%
{\displaystyle\int\limits_{0}^{2R}}
\frac{\left[  1-h_{\frac{3}{2}}\left(  -\frac{4\zeta^{2}}{\pi^{2}\omega^{2}%
}\right)  \right]  }{(R^{2}-\zeta^{2}/4)^{%
\frac12
}}\frac{\zeta^{2}d\zeta}{4R}=\frac{\pi R}{2}\left[  \allowbreak1-h_{2}\left(
-\frac{16R^{2}}{\pi^{2}\omega^{2}}\right)  \right]  \label{DISK}%
\end{equation}
The last but one argument of the hypergeometric function is used as a
subscript in our notation $h_{2}(x)$ $\equiv$ $_{2}\digamma_{3}[\{\frac{1}%
{3},\frac{3}{4}\},\{\frac{1}{2},1,2\},x]$.

\section{3.8. Lineshape for spherical particles of the radius $R$}

In the third row (c) and in rows below (d,e,f) of Fig.10, the scattering
matter consist of 0D-objects, i.e., spherical particles. In (c) all the
particles have spherical shape and constant radius. The RTD function coincides
with $d(\zeta)$ when the particles are distributed over the plane surface $S$
in such a way that shadowing of one particle by another is excluded. Again, we
have the impact distance $b=\sqrt{R^{2}-\zeta^{2}/4}$, however, $dS/S$ is
now\ $(\pi R^{2})^{-1}(2\pi bdb/d\zeta)=\zeta/2R^{2}(0<\zeta<2R)$. Then%

\begin{equation}%
{\displaystyle\int\limits_{0}^{2R}}
\left[  1-h_{\frac{3}{2}}\left(  -\frac{4\zeta^{2}}{\pi^{2}\omega^{2}}\right)
\right]  \frac{\zeta^{2}d\zeta}{2R^{2}}=\frac{4R}{3}\left[  \allowbreak
1-h_{\frac{5}{2}}\left(  -\frac{16R^{2}}{\pi^{2}\omega^{2}}\right)  \right]
\label{Ball}%
\end{equation}
Similarly to Eqs.(\ref{FT},\ref{DISK}) the last but one argument is used as a
subscript $h_{\frac{5}{2}}(x)\equiv$ $_{2}\digamma_{3}[\{\frac{1}{3},\frac
{3}{4}\},\{\frac{1}{2},1,\frac{5}{2}\},x]$.

\section{3.9. Lineshapes generated by several characteristic particle size
distributions}

Assuming the spherical particles to have some distribution in size we observe
(Fig. 10 (d, e) that the PSD's of rectangular and triangular shape result in
the lineshapes with smeared shoulders. No shoulders can be apperceived for the
cases (f) and (g), related to exponential PSD and exponential CLD,
respectively. All the PSD's except (g) are normalized to have the same area
and the same mean particle size $\langle r\rangle=R$. The CLD's $d(\zeta)$ are
derived from PSD's $F(r)$ using the formula\cite{Olson}:%

\begin{equation}
d(\zeta)=\frac{\zeta}{2R^{2}}%
{\displaystyle\int\limits_{\zeta/2}^{\infty}}
F(r)dr \label{CLD}%
\end{equation}

In case of exponential PSD (f), i.e., $F(r)=(1/R)\exp(-r/R)$, the
Eq.(\ref{CLD}) results in the CLD $d(\zeta)=(\zeta/4R^{2})\exp(-\zeta/2R)$.
Integrating this $d(\zeta)$ with the right-hand side of the Eq. (\ref{FT}) and
$\zeta$ from the Eq. (\ref{rh}) we obtain the spectral lineshape
\begin{align}
&
{\displaystyle\int\limits_{0}^{\infty}}
\left[  1-h_{\frac{3}{2}}\left(  -\frac{4\zeta^{2}}{\pi^{2}\omega^{2}}\right)
\right]  \frac{\zeta^{2}\exp(-\zeta/2R)d\zeta}{4R^{2}}=\nonumber\\
&  =4R\allowbreak\left[  1+\frac{2\sqrt{\pi}R\sin(3\Phi/2)}{\omega(\pi
^{2}+\frac{64R^{2}}{\omega^{2}})^{%
\frac34
}}-\frac{\sqrt{\pi}\cos(\Phi/2)}{(\pi^{2}+\frac{64R^{2}}{\omega^{2}})^{%
\frac14
}}\right] \label{ExpPSD}\\
& \nonumber
\end{align}
Here $\Phi=\arctan(8R/\pi\omega)$. As far as the relative weight of smaller
particles increases from top to bottom of Fig. 10 the shoulders became less
articulated and disappear in (f).

Further increase in the fraction of small particles is the case (g) related to
divergence of $F(r)$ at $r\longrightarrow0$. Substituting $F(r)=(r/R^{2}%
+2/r)\exp(-2r/R)$ into Eq. \ref{CLD} we obtain the simple exponential CLD
$d(\zeta)=R^{-1}\exp(-\zeta/2R)$, which is broadly found among the natural
particulate materials. Since this $F(r)$ has a singularity at
$r\longrightarrow0$ , such a PSD cannot be normalized in the same way as the
PSD's (d), (e) and (f). However, $d(\zeta)$ can be normalized in a usual way $%
{\textstyle\int}
d(r)dr=1$ with the average chord length $\langle\zeta\rangle=R$. Many kinds of
complex interfacial media exhibit the exponential CLD. Numerous examples of
the exponential CLD's were demonstrated \cite{Levitz} among the porous
borosilicate glasses, heterogeneous catalysts, cements etc. Also, if we mix
the hard spherical particles of non-resonant media with soft resonant
background we expect an exponential CLD for the resonant
component\cite{Olson2008}. The absorbers for NFS are frequently prepared
through mixing the resonant nanoparticles (e.g., $^{57}$Fe$_{2}$O$_{3}$ in
\cite{Rykov}) with larger particles of non-resonant filling medium.

The divergence of $F(r)$ at small particle radius ($r\longrightarrow0$) is so
strong that the integral $%
{\textstyle\int}
F(r)dr$ is also divergent. This nonitegrability implies that the spherical
shape assumption (Eq. \ref{CLD}) is rather unphysical for the exponential CLD
$d(\zeta)=(1/R)\exp(-\zeta/R)$ \cite{Olson}. Indeed, this $d(\zeta)$ is the
chord length distribution in the background material filling the interstices
between spheres\cite{Olson2008}. Integrating this $d(\zeta)$ with the
right-hand side of the Eq. (\ref{FT}) and $\zeta$ from the Eq. (\ref{rh}) we
obtain the lineshape of such a background:%

\begin{equation}%
{\displaystyle\int\limits_{0}^{\infty}}
\left[  1-h_{\frac{3}{2}}\left(  -\frac{4\zeta^{2}}{\pi^{2}\omega^{2}}\right)
\right]  \frac{e^{-\frac{\zeta}{R}}\zeta d\zeta}{R}=R\left[  \allowbreak
1-\frac{\sqrt{\pi}\cos(\phi/2)}{(\pi^{2}+\frac{16R^{2}}{\omega^{2}})^{%
\frac14
}}\right]  \label{exp}%
\end{equation}
Here $\phi=\arctan(4R/\pi\omega).$

\section{3.10. Determination of the particle size distribution parameters}

The Eqs. (\ref{FT}), (\ref{Ball}), (\ref{ExpPSD}) and (\ref{exp}) were
employed to fit the upper-frequency line from the $\exp(\tau)$-sharpened
spectrum of Fig. 8. First, assuming a layer of uniform thickness the
Eq.(\ref{FT}) was used. Fitting the line at 185 MHz with the hypergeometric
function $1-h_{3/2}\left(  -4\zeta^{2}/\pi^{2}\omega^{2}\right)  $ has
resulted in the value of the dimensionless thickness $\zeta_{0}=4.36$
(Fig.11). According to the Eq.(\ref{ch}) the total Mossbauer thickness is
$\zeta_{0}^{\text{tot}}=4\zeta_{0}\simeq17.4$. This is very close to the value
of $\zeta^{\text{tot}}=19.5$ obtained above in time domain through fitting the
NFS spectrum as a whole.%
\begin{figure}
[ptb]
\begin{center}
\includegraphics[
height=2.6351in,
width=3.467in
]%
{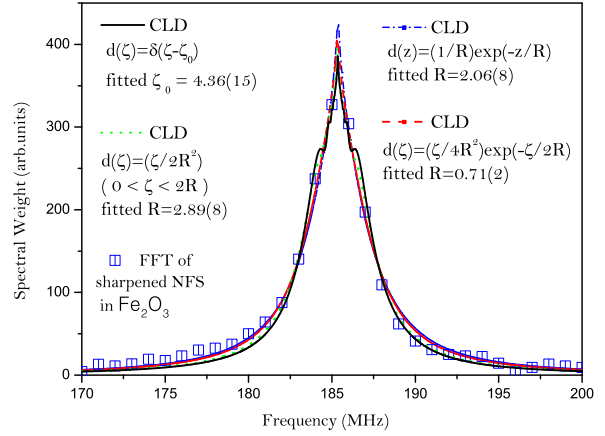}%
\caption{The section from 170 MHz to 200 MHz of the quantum beat frequency
spectra (square-shaped symbols) fitted with the right-hand expressions in Eqs.
(\ref{ExpPSD}) and (\ref{exp}) (lines). }%
\end{center}
\end{figure}

Turning from dimensionless $\zeta^{\text{tot}}$ to metric units of
$z=\zeta^{\text{tot}}/\pi\mu^{\text{tot}}$ and using $\mu^{\text{tot}}%
=\sigma_{0}\rho f_{\text{LM}}\eta$ with $\sigma_{0}=2.56\cdot10^{-18}$
cm$^{2}$, $\rho=3.96\cdot10^{22}$ cm$^{-3}$, $f_{\text{LM}}=0.8$,
$\eta=0.0212$, we obtain the total coefficient of nuclear absorption for
unenriched hematite $\mu^{\text{tot}}=0.172$ $\mu$m$^{-1}$. From the
Eq.(\ref{ch}) the layer thickness $z_{0}$ that corresponds to $\zeta
_{0}^{\text{tot}}=17.4$ is (shown by the position of $\delta$-function in Fig.12):%

\begin{equation}
z_{0}=4\zeta_{0}^{\text{tot}}/\pi\mu^{\text{tot}}=16\zeta_{0}/\pi
\mu^{\text{tot}}=129\mu m \label{uni}%
\end{equation}
%

\begin{figure}
[ptb]
\begin{center}
\includegraphics[
height=2.3155in,
width=3.4429in
]%
{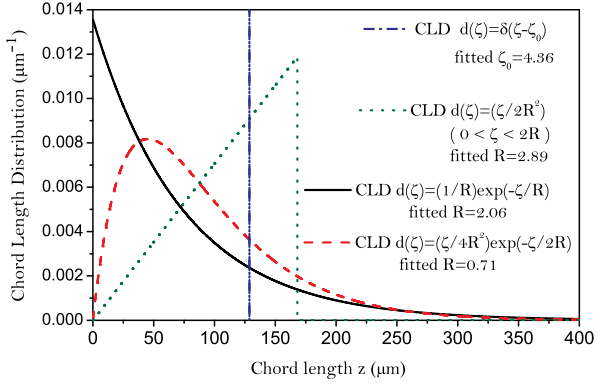}%
\caption{Chord length distributions $d(z)$ used to fit the highest-frequency
line of the NFS Fourier spectra of Fig.8 at the values of the dimensionless
parameters: $\zeta_{0}=4.36$ in $d(\zeta)=\delta(\zeta-\zeta_{0})$, $R=2.89$
in $d(\zeta)=\frac{\zeta}{2R^{2}}$, $R=2.38$ in $d(\zeta)=\frac{1}{R}%
e^{-\frac{\zeta}{R}}$ and $R=0.82$ in $d(\zeta)=\frac{\zeta}{4R^{2}}%
e^{-\frac{\zeta}{2R}}$.}%
\end{center}
\end{figure}

Certainly, a sample formed by the particles of Fe$_{2}$O$_{3}$ of a finite
size appears worthy of a better characterization. Assuming all the particles
having the same size and spheric shape, we can estimate the radius of particle
using Eq.(\ref{Ball}). The fit results in the value of $R$ of $2.84$. In this
model, the thickness is distributed linearly between $0$ and $2R$ (Fig.12).
The particle diameter of 168 $\mu$m is larger by 30\% than uniform thickness
$z_{0}$ from the Eq.(\ref{uni}). Here again as in Eq.(\ref{uni}) the
hematite-specific length factor $r_{0}=16/\pi\mu^{\text{tot}}=29.61\mu$m
brings the diameter $2Rr_{0}$ to the metric scale. The goodness of fit becomes
better when we proceed from uniform layer to the ball-shaped particles,
namely, the coefficient of determination $\mathbf{R}^{\mathrm{2}}$ increases
from 0.991 to 0.994. Although the increase is not large, the usage of the
$\mathbf{R}^{\mathrm{2}}-$coefficient is justified at comparing the goodness
of fit for different models because all of our models use the same number of parameters.

Among the models with distributions of particle sizes two models were tested,
according to the Eqs.(\ref{ExpPSD},\ref{exp}). Goodness of fit was slightly
better for the first (Eq. \ref{ExpPSD}, $\mathbf{R}^{\mathrm{2}}=0.995$) than
for the second (Eq. \ref{exp}, $\mathbf{R}^{\mathrm{2}}=0.993$). Corresponding
chord lengths distributions are also shown in Fig.12. The quality of fit
between these two distributions is not much different, the values of fitted
radius $R$ for the CLD of spheres is nearly 3 times smaller than the value of
$R$ for the CLD of the background between spheres. In the chord length range
$38$ $\mu$m $<z<260$ $\mu$m the CLD density for spheres is larger than the CLD
density for the background. Opposite is true elsewhere. Clearly, the excess in
the range $38$ $\mu$m $<z<260$ $\mu$m is compensated by the contribution of
large particles, although their CLD density above $260$ $\mu$m\ is below 5\%.
In this sense, the chord length of $260$ $\mu$m is the characteristic length
invariant of the fitting model employed. Interestingly, this "inhomogeneous"
length is twice larger than the value obtained in Eq.(\ref{uni}) for the
homogeneous layer.

\section{\protect\linebreak4. Conclusions}

Starting from several characteristic states of scattering matter (foil,
filaments, particles) we have solved the problem of determination of
corresponding spectral lineshape parameters for the nuclear forward scattering
Fourier spectra. The method is useful in the range of thicknesses where
quantum beat envelopes vary slowly compared to the quantum beats themselves.
In the frequency domain, we are able to predict the lineshapes determined by
Fourier transforms of the envelopes. These lineshapes are of polaritonic
nature since they originate from the energy exchange between the radiation
field and nuclear excitation. Both the radiation and the nuclear currents are
the components of the compound quasiparticle termed the nuclear exciton
polariton. From these lineshapes we derive the parameters of resonant
thickness distribution.

When fragments of resonant and nonresonant media are interleaved so that the
standard methods of neutron or x-ray small-angle scattering (SAS) cannot
distinguish between fragments of similar electronic or nuclear density, the
proposed technique would be able to resolve between them. Therefore, if a
fraction of the $^{57}$Fe-containing 'particles' cannot be detached from a
'membrane', then the methods of SAS would produce the information on the
density distribution in the system 'particles+membrane'. Separately, the
particles subsystem can be studied by NFS without membrane detaching that is
most frequently unrealizable. Supported catalysts present the archetype of
future applications.

Practical applications of the proposed analysis are quite feasible already at
the existing beamlines of the synchrotron rings of third generation. The
sampling frequency does not require any improvements, because it can be needed
only for very thick samples. The time resolution of the avalanche photodiode
detectors ($\sim$0.1 ns) correspond closely to the currently applied sampling
frequency. Much more important for the proposed technique would be the
improvements in the high-rate characteristics of the detectors and in the span
of the time window between electron bunches. Change from multiple-bunch to
single-bunch regime is crucial to achieve a better cutoff time $\tau_{\max}$.
Also, with the advent of novel avalanche detectors and detector arrays
\cite{Kishimoto2003, Kishimoto} the region of missing data at small $\tau$
might be shortened.

Inverse problem of finding the characteristic parameters of the particle chord
length distribution from the polaritonic lineshapes would be possible to solve
using regularization methods. Then, not only PSD parameters could be refined
starting from one or another hypothesis, but also the full PSD profiles could
be reconstructed.

This work was supported by RFBR-JSPS Grant 07-02-91201.


\bigskip

\begin{thebibliography}{99}                                                                                               %


\bibitem {Kohn1}\Name{Kagan Yu., Afanas'ev  A.M., \and Kohn V.G.} \REVIEW{J.
Phys.} {C12}{1979}{615}.

\bibitem {Svergun}\Name{Feigin L.A. \and Svergun D.I.} \Book{Structural
Analysis by Small-Angle X-ray and Neutron Scattering } \Editor{George W.
Taylor} \Publ{Plenum Press, New York} \Year{1987} \Page{44}.

\bibitem {Bausk}\Name{Bausk N.V., Erenburg S.B. , \and Mazalov L.N.} \REVIEW{J. Synchrotron Rad. } {6}{1999}{ 268-270}.

\bibitem {Gille}\Name{ Gille W.} \REVIEW{Eur. Phys. J. B.} {17}{2000}{
371-383}.

\bibitem {Gerdau}\Name{ Gerdau E., R\"uffer R., Hollatz R., and Hannon J. P.}
\REVIEW{Phys. Rev. Lett.} {57}{1986}{
1141}.

\bibitem {Hastings}%
\Name{Hastings J.B., Siddons D.P., van B\"urck U., Hollatz R.,  \and Bergmann U.}
\REVIEW{Phys. Rev. Lett.
}{66}{1991} {770}.

\bibitem {Buerck}%
\Name{van B\"urck U., Siddons D.P., Hastings J.B., Bergmann U. \and  Hollatz R.}
\REVIEW{Phys. Rev. B.
}{46}{1992} {6207}.

\bibitem {Smirnov}\Name{Smirnov G. V.} \REVIEW{Hyperfine Interact.} {97/98}{1996}{551}.

\bibitem {BuerckHyp}\Name{van B\"urck U.} \REVIEW{Hyperfine Interact.} {123/124}{1999}{483-509}.

\bibitem {Shvyd'ko}\Name{Shvyd'ko Yu. V. , van B\"urck U. ,Potzel  W.
,Schindelmann  P.,Gerdau  E.,Leupold  O.,Metge  J.,R\"uter  H.D., and
Smirnov G.V.} \REVIEW{, Phys. Rev. B} {57}{1998}{ 3552-3561}

\bibitem {Rykov}\Name{Rykov A.I., Rykov I. A., Nomura K., \and Zhang X.} \REVIEW{Hyperfine Interact.} {163}{2005}{29-56}.

\bibitem {Smir2}\Name{Smirnov G. V.} \REVIEW{Hyperfine Interact.} {123/124}{1999}{31-77}.

\bibitem {SB}\Name{Shvyd'ko Yu. V. , van B\"urck U.} \REVIEW{Hyperfine Interact.} {123/124}{1999}{511-527}.

\bibitem {Sm2007}%
\Name{Smirnov G.V., van B\"urck U., Arthur G., Brown G.S., Chumakov A.I., Baron A.Q.R., Petry W., \and Ruby S.L.}
\REVIEW{Phys. Rev.
A}{76}{2007} {043811}..

\bibitem {Kohn2}\Name{Kohn V.G.  \and Smirnov G.V.} \REVIEW{Phys. Rev.
B}{76}{2007} {104438.}.

\bibitem {Olson}\Name{Olson G.L., Miller D.S., Larsen E.W. \and Morel J.E.} \REVIEW{J. Quantitative Specroscopy and Radiative Transfer} {101}{2006}{269-283}.

\bibitem {Levitz}\Name{Levitz P., \and Tchoubar D.} \REVIEW{J. Phys. I
France} {2}{1992}{771-790}.

\bibitem {Olson2008}\Name{Olson G.L.} \REVIEW{Annals Nucl. Energy} {35}{2008}{2150-2155}.

\bibitem {Kishimoto2003}\Name{ Kishimoto S., Yoda Y., Seto S., Kitao S.,
Kobayashi Y., Haruki R. \and Harami T.} \REVIEW{Nuclear Instruments and
Methods in Physics Research A} {513}{2003}{193}.

\bibitem {Kishimoto}\Name{ Baron A.Q.R., Kishimoto S., Morse J., \and Rigal
J.-M.} \REVIEW{J. Synchrotron Rad.} {13}{2006}{ 131-142}.
\end{thebibliography}
\end{document}